%Paper: gr-qc/9403016
%From: JENNIE TRASCHEN <lboo@phast.umass.edu>
%Date: Wed, 09 Mar 1994 12:30:39 -0500

\input harvmac
\sequentialequations
\noblackbox
\def\bk{\hfill\break}
\def\k{\kappa}
\def\lam2{\lambda ^2}
\def\at{A_t}
\def\pary{\partial _y}
\def\part{\partial _t}
\def\pho{\phi _o}
\def\wep{(\omega - e\phi _o )}
\def\ra{\rightarrow}
\def\w-e{(\omega -e )}
\def\crit{\sqrt{{2M m^2 \over e}} [(1- {e^2 \over m^ 2 })\Delta +\epsilon
]^{1\over 2} }
\def\dt{\Delta t}
\def\dy{\Delta y}

\lref\chop{M. Choptuik,
{\it Universality and Scaling in Gravitational Collapse of a
Massless Scalar Field},  Phys.
Rev.Lett. {\bf 70}, 9 (1993).}
\lref\evans{ A. Abrahams and C. Evans,{\it Critical Behavior and Scaling in
Vacuum
Axisymmetric Gravitational Collapse},  Phys.Rev.Lett. {\bf 70}, 2980, (1993).}
\lref\jtrf{J.Traschen and R.Ferrell, {\it Quantum Mechanical Scattering
of Charged Black
Holes}, Phys. Rev. {\bf D45}, 2628 (1992).}
\lref\shir{ K.Shiraishi, {\it Classical and Quantum scattering of Maximally
Charged Black
Holes}, Int.J.Mod.Phys. {\bf D}, 59 (1993).}

\noblackbox
\sequentialequations
%%%%%%%%%%%%%%%%%%%%%%%%%%%%%%%%%%%%%%%%%%%%%%%%
% Macros to remove Harvard titlepage from harvmac
%                   S.B.G. 3/91
%

%
%
%%%%%%%%%%%%%%%%%%%%%%%%%%%%%%%%%%%%%%%%%%%%%%%%%

\def\Title#1#2{\ifx\answ\bigans \nopagenumbers
\abstractfont\hsize=\hstitle\rightline{#1}%
\vskip .5in\centerline{\titlefont #2}\abstractfont\vskip .5in\pageno=0
\else \rightline{#1}
%\abstractfont %\hsize=\hstitle
%\rightline{#1}%
\vskip .8in\centerline{\titlefont #2}%\abstractfont
\vskip .5in\pageno=1\fi}
\ifx\answ\bigans

scaled\magstep3
\else

scaled\magstep3
 
 \font\absi=cmmi10 scaled\magstep1
\font\absis=cmmi7 scaled\magstep1 \font\absiss=cmmi5 scaled\magstep1
\font\abssy=cmsy10 scaled\magstep1 \font\abssys=cmsy7 scaled\magstep1
\font\abssyss=cmsy5 scaled\magstep1 
\skewchar\absi='177 \skewchar\absis='177 \skewchar\absiss='177
\skewchar\abssy='60 \skewchar\abssys='60 \skewchar\abssyss='60
\fi

%%%%%%%%%%%%%%%%%%%%%%%%%%%%%%%%%%%%%%%%%
%
%
% Title Page
%
%
%%%%%%%%%%%%%%%%%%%%%%%%%%%%%%%%%%%%%%%%%
\Title{\vbox{\baselineskip12pt
\hbox{UMHEP-405}
\hbox{gr-qc/9403016}}}
{\vbox{\centerline{Discrete Self-Similiarity and Critical Point Behavior }
\vskip 0.2in
\centerline{in Fluctuations About Extremal Black Holes}}}
\baselineskip=12pt
\centerline{Jennie Traschen}
\bigskip
\centerline{\sl Department of Physics and Astronomy}
\centerline{\sl University of Massachusetts}
\centerline{\sl Amherst, MA 01003-4525}
\centerline{\it lboo@phast.umass.edu}
\bigskip
\centerline{\bf Abstract}

The issues of scaling symmetry and critical point behavior are studied for
fluctuations
about extremal charged black holes. We consider the scattering and capture of
the spherically
symmetric mode of  a charged, massive test field on the background spacetime of
a black hole with charge $Q$ and mass $M$. The spacetime geometry near the
horizon of
a $|Q|=M$ black hole has a scaling symmetry, which is absent if $|Q|<M$, a
scale being introduced
by the surface gravity.  We show that this symmetry leads to the existence of a
self-similiar solution for the charged field near the horizon, and further,
that there
is a one parameter family of discretely self-similiar solutions . The scaling
symmetry,
or lack thereof, also shows up in correlation length scales, defined in terms
of the
rate at which the influence of an external source coupled to the field dies
off.
It is shown by constructing the Greens functions,
that an external source has a long range influence on the extremal
background, compared to a correlation length scale which falls off
exponentially fast
in the $|Q|<M$ case. Finally it is shown that in the limit of $\Delta \equiv
(1-{Q^2 \over
M^2})^{1\over 2} \rightarrow 0$ in the background
spacetime, that infinitesimal changes in
the black hole area vary like $\Delta ^{1\over 2}$.

\Date{3/94}

%%%%%%%%%%%%%%%%%%%%%%%%%%%%%%%%%%%%%%%%%%%%%%%%%%%%%%%%%%%%%%%%%%%%%%%%%%%%%%
\newsec{Introduction}

Recently, numerical studies of gravitational collapse, have shown
scaling and critical point type  behavior in the formation of zero mass,
neutral blackholes
\chop , \evans . These studies showed several interesting properties; in the
zero
mass limit, the wave form of the collapsing wave always evolved to a particular
form, which was discretely self-similiar in appropriate time and space
variables.  The mass
of the formed black hole had non-analytic  dependence on a variety of
parameters
measuring the difference in the strength of the wave from some critical value,
and
the exponent was found to be universal.

Suppose that  charged particles collapse to form a black hole. In this case,
the black hole
must have a mass which is greater than or equal to its charge.
Would critical point type behavior be seen in fluctuations about the minimal
area?
Or more generally,
would such behavior be seen in the interactions between charged wave packets
and an already existing charged  black hole, in the extremal limit?

There are two geometrical reasons why this might occur. First,
the spacetime geometry near the horizon of a $Q=M$ black hole has an infinite
throat
(the metric approaches a Robinson-Bertotti metric).
The throat has no scale, and the metric has a dilatation symmetry, which means
that test fields on this background will have a scale invariance.  By contrast,
the geometry near the horizon of a non-extremal
black hole has a scale set by the surface gravity $\k$.  Second, dust with mass
density
equal to its charge density can be placed in arbitary configurations and will
stay in
equilibrium with other configurations of such  dust , and  with arbitrary
distributions of
charge equal to mass black holes. There is no particular size  of charge equal
to mass
dust that is needed for a force balance.
This is reminiscent of the picture
of fluctuations on all length scales occuring at a critical point.

In this paper we will focus on the questions of scaling invariance and
self-similiar
solutions, correlation length scales, and  how  fluctuations in the area of the
black hole depend on $\Delta \equiv \sqrt{1-{Q^2 \over M^2 }}$, as $\Delta \ra
0$. Of
course one would like to have exact solutions describing wave packets of
charged
fields scattering off a charged black hole, analogous to the numerical work
\chop ,\evans . Here,
in order to make some progress analytically, we will study a charged, massive,
test  field
scattering off a fixed black hole background with charge $Q$ and mass $M$. This
is a consistent approach
since (1) we are interested in the limit where the change in the area is
infinitesimal,
and (2) because there is already a black hole present to do perturbation theory
around,
unlike the neutral black hole case.
We imagine an initial wavepacket
which heads towards the black hole, part of which is scattered and part of
which
is captured. One is interested in the form of the captured wave, in particular
to see if it shows scaling behavior when the background spacetime approaches
extremality.  We will first show that near the horizon of an extremal black
hole,
the wave does have a scaling symmetry
when the background spacetime is extremal. Amongst these solutions there is one
which
is self-similiar, and has a translation invariance in logarithmic time and
logarithmic
radial coordinates. Further, we will show that near the horizon, there exists a
set of eigenfunctions
of the wave equation which each of which has a discrete self-similiarity.

Second, addressing the issue of correlation lengths is a bit confusing--
this is not (at least apparently) a statistical system with degrees of freedom
to average
over. However, one can construct the Greens functions for the wave equations,
and these
tell what the response of the test field is to a source.  The Greens functions
show
that on the extremal background the influence of a source is long range,
whereas
on a background with $Q<M$, it falls off exponentially fast. Thirdly,
the captured part of the wave adds mass and charge to the black hole.
We will show that the resulting change in the area of the black hole goes like
$\Delta ^{1\over 2}$ in the limit where $\Delta \ra 0$.

To avoid repeated absolute value signs, we will take the charge of the black
hole
to be positive. We will use units with $G=1$.

\newsec{Description of the System}
Consider a charged scalar field on a Reissner-Nordstrom background spacetime,
with metric given by
\eqn\metric{ds^2 =\lam2 (-dt^2 +dy^2 ) +R^2 d\Omega ^2,\qquad
\lam2 =1-{2M\over R} +{Q^2 \over R^2}}
Here $y$ is the usual tortoise coordinate, $dy=dR/\lam2 $,
$R=R(y)$ and $Q$ an $M$ are the electric charge and mass of the spacetime.
Assume that the charge $Q$ is positive.
Choose the gauge of the electrostatic potential such that it vanishes on
the horizon, $R=R_H$,
\eqn\gauge{\at =\pho -{Q\over R} \ , \  \  \pho\equiv
{Q\over R_H}}
The quantity $\pho$ is the then the difference in the electrostatic potential
between the horizon and infinity. In the context of the first law of black
hole thermodynamics, $\pho$ is conjugate to the electric charge $Q$.
Note that for $Q=M$,
\eqn\extremal{\at =\lambda = 1-{M\over R},\qquad \pho =1.}

The matter action, for a scalar field of mass $m$ and charge $e$, is given by
\eqn\action{S_m =\int \sqrt{-g}(D_a\Phi D^a \Phi ^* -m^2 \Phi \Phi ^*),}
where the gauge covariant derivative $D_a \Phi =(\partial _a -ie A_a )\Phi $.
The equation of motion for the scalar field is
\eqn\eomone{ \nabla _a \nabla ^a \Phi -2ieA_a \nabla ^a \Phi -(m^2 +e^2 A_a A^a
)\Phi =0 }
We will consider spherically symmetric waves on the background \metric .
Let $\psi =R(y)\Phi$. Then the equation of motion for the field \eomone\
becomes
\eqn\eom{\pary ^2 \psi -\part ^2 \psi +2ie\at \part \psi +(e^2 \at ^2
-m^2 \lam2  -V_{grav})\psi =0,}
with the potential $V_{grav}$ given by
\eqn\potential{V_{grav} ={\pary ^2 R \over R}
={2M\over R^3 }\lam2 (1-{Q^2 \over MR } )}

Lastly, we summarize the
behavior of the system near the horizon. In the tortoise coordinates,
the black hole horizon,
$R=R_H$, is at $y\rightarrow -\infty $. One finds near the horizon that:\bk
for $Q=M$,
\eqn\limit{ R-M \ra -{M^2 \over y},\ {\rm and}\  \at =\lambda \ra -{M\over y},}
while for $Q<M,$
\eqn\nonext{R-R_H \ra R_H e^{2\k y} ,\ \lam2 \ra 2\k R_H e^{2\k y} ,
\ \at \ra\pho e^{2\k y},}
where $\k ={1\over 2}\partial _R \lam2 |_{R_H}$ is the surface gravity at the
horizon , which vanishes for the extremal black hole.

\newsec{Scaling Behavior  and Self-Similiar Solutions near the Horizon of
Extremal Black Holes}
{}From the asymptotic behaviors of the gauge potential and metric function, it
follows that the wave equation \eom \ for the scalar field is invariant under
the rescaling
\eqn\rescale{y\rightarrow ay ,\ t\rightarrow at ,\ a=constant . }
This implies that there are solutions to \eom\ which are functions only of the
ratio $t/y$. Equivalently, in term of logarithmic coordinates $\bar{t}=\ln t
,\bar{y} =\ln (-y )$, there are solutions of the form $\psi =F(\bar{t} -
\bar{y})$. These solutions have the translation invariance,
corresponding to self-similiarity in the original $t,y$ coordinates, $\psi
(\bar{t} +D, \bar{y} +D) =\psi (\bar{t} ,\bar{y})$ for any $D$. By contrast,
the
field equation  on the nonextremal black hole spacetime does not have a scaling
invariance--the surface gravity $\k$ introduces a scale.

The scaling  invariance is a reflection of an additional dilatation symmetry of
the metric
near the horizon, $ds^2  \rightarrow {M^2 \over y^2}( -dt^2 +dy^2) +M^2 d\Omega
^2$,
with the dilatation killing vector  $\xi ^a =t({\partial \over \partial t})^a
+y({\partial
\over \partial y})^a$.
The symmetry is actually best seen in a slightly different set of coordinates,
in which the metric function is used as a coordinate. Let $x =1-{M\over R}$.
Then the wave equation  \eomone \ becomes
\eqn\scaleq{
-\part ^2 \Phi +2iex\part \Phi +x^2 {(1-x)^4 \over M^2 } \partial _x
(x^2 \partial _x \Phi ) +x^2 (e^2 -m^2 )\Phi =0. }
As $x\ra 0$ the horizon is approached and
the differential equation becomes invariant under transformations of the form
$x\ra {1\over a}x, t\ra at$, with $a=constant$. The utility of the coordinate
$x$ is that it is consistent to include the term $(e^2 -m^2 )\Phi $ .
In the tortoise coordinates above it is not clear that one can legitimately
retain this term, while ignoring higher order terms in the inversion
relation between $y$ and $R$.

Now let $w=xt$ and look for solutions  of the form
\eqn\discrete{ \Phi (x,t) = t^{i\nu}F_{\nu} (w) }
In terms of the logarithmic time $\bar{t}$ introduced above,
the prefactor is $e^{i\nu {\bar t}}$,
and so the solution will have a discrete self-similiarity ${\bar t}\ra {\bar t}
+
{2\pi \over \nu }$ and ${\bar x}=-ln x \ra {\bar x} +{2\pi \over \nu}$
For the eigenvalue $\nu =0$ this is the  continuously
self-similiar solution discussed above, depending only on $w$.

The wave equation \scaleq\ becomes, for $x\ll 1$,
\eqn\besteq{ (1-{w^2 \over M^2}) F_{\nu }'' (w) +2(-ie -{w\over M^2 }
+{i\nu\over w})
F_{\nu}'  +(m^2 -e^2 +{2e\nu\over w} -{\nu ^2 +i\nu \over w^2 }) F_{\nu} =0   }
Analysing the solutions to this equation is a topic for future study, however,
we have
shown that the wave equation for a massive charged field on the background
spacetime of an  extremal black hole, has discretely self-similiar solutions,
as the
horizon is approached. It is worth noting that for $\nu =0$ the differential
equation
becomes
\eqn\niceq {{d\over dw} \left( (1-{w^2 \over M^2 })f'(w) -2ief \right)  -(e^2
-m^2
)f=0. }
For $e=m$ it is simple to find the solution,
\eqn\nicesol{ f(w) =C_1 +C_2 ({M+w
\over M-w})^{ieM}. }
In terms of the tortoise coordinate $y$ this is
\eqn\nicersol{\Phi =C_1 +C_2 ({1-t/y \over 1+t/y })^{ieM} ,\ as\
y\ra -\infty.  }

A general solution to the wave equation near the horizon can be written as a
sum
of the eigenfunctions \discrete . However, an arbitrary sum will no longer be
discretely self-similiar. So,
perhaps the most interesting question is, do generic wave packets starting
in the flat region (or ANY packets for that matter), evolve into a packet which
is a special sum of the modes \discrete , \ such  that the sum is discretely
(or
continuously ) self-similiar?
We do not yet know the  answer to this, but one can imagine at least two ways
in which this could happen. When the eigenvalue problem is solved with
appropriate
boundary conditions for a wave packet incoming from the flat region, it may
be that the eigenvalues $\nu$ are actually  quantized , say $\nu _n = n {C\over
M}$. Then the solution would have a discrete self-similiarity with $D ={2\pi
M\over C}$. Alternatively,
it could be that the imaginary part of the frequency $\nu $ is positive, so
that
the lowest frequency dominates at late times. Or, it may be that the evolution
from
wave packets ``at infinity'' is not self-similiar! It will certainly be of
interest to resolve
this question.

\newsec{Scattering and Asymptotic Solutions}
The scaling symmetry can be displayed in terms of the greens function for the
scalar field equation.  The greens function describes  how the field propagates
in
response to an external source. We will show that the greens function for the
$Q=M$
case has long range $1/y$ correlations, compared to the greens function for the
$Q<M$ case, in which correlations die off exponentially with scale $\k ^{-1}$.
To this end, we will find eigenmodes of the wave equation in the asymptotic
regimes,
and use the solutions to construct the greens functions. Further,
the scattering behavior  of the eigenmodes will be anlyzed, to determine what
scatters and what is captured.
{}From this, the change in the horizon area $\delta A$  will be found,
when a small amount of mass and charge is captured.  $\delta A$ has
non-analytic
behavior as extremality is approached, whereas away from $Q=M$,  $\delta A$ is
linear in the added mass and charge.

To put the wave equation in the form of a scattering problem,
first, fourier transform in time; let
\eqn\time{\part \psi =-i\wep \psi}
Then \eom \ becomes
\eqn\wave{\pary ^2 \psi +[ \wep ^2 +2e\wep \at +e^2\at ^2  -m^2 \lam2 -V_{grav}
] \psi =0}
or,
\eqn\wavetwo{\pary ^2 \psi +[ k^2 -V -V_{grav} ] \psi =0 \  ,\  k^2 +m^2 \equiv
\omega ^2 }
where
\eqn\limitpot{V\rightarrow {2\over R}(e\omega Q -m^2 M )  -(e^2 -m^2 )
{Q^2 \over M^2 } ,\ R\rightarrow \infty }
At large distances  $R\ra y$ and the form of the potential $V$ is the same in
all cases.
$V_{grav}$ falls off like $R^{-3}$. However,  near the horizon there is a
qualitative
difference for the extremal and non-extremal cases: \bk
For $Q<M$,
\eqn\nonextpot{
V\rightarrow 2\omega e\pho -e^2 \pho ^2 -m^2  -(2e\pho \wep  -m^2 2\k R_H )
e^{2\k
y}+ O(e^{4\k y}) \ , \ y\rightarrow -\infty .}
For $Q=M$,
\eqn\extpot{V\rightarrow -m^2 -e^2 +2e\omega +2e(\omega -e ){M\over  y} -(e^2
-m^2 ){M^2 \over y^2 },
\  y\rightarrow  -\infty .}

Analysis of the potentials shows the following qualitative features of the
scattering
problem:  At large $y$, the potential falls off quite slowly (like $y^{-1}$). A
WKB approximation,
shows that the transmission is exponentially suppressed if the incident wave is
under the barrier.
For the purposes of the following discussion then, it will be sufficient to
approximate
the capture cross section as a step--if the wave is over the barrier, it is
captured, and
if it is under the barrier, the wave is scattered. Essentially, we are working
in a geometrics
optics approximation. (However, we know that the approximation is a good one
here,
because the scattering problem is similiar to that in \jtrf, \shir , in which
the scattering is
worked out in analytic and numerical detail.)

So what we need is the criterion for an eigenmode to be over the barrier. Now,
this is not quite a standard scattering equation, because the height
of the potential depends on the incident wave frequency $\omega$. But studying
the
potential, one finds that to be over the barrier, a wave with frequency
$\omega$ must satisfy
\eqn\capture{\wep > {m^2 \over e}{\Delta +\epsilon \over \sqrt{1-\Delta ^2}}
\    \  ,\epsilon >0}
\eqn\param{where \ \  \Delta \equiv \sqrt{1-{Q^2 \over M^2}} .}
$\Delta$ is a parameter which measures how close the spacetime is to extremal.
$\epsilon$ is any number greater than zero, and merely insures that  a wave
packet
centered on the frequency $\omega$ reaches the horizon in finite time.
This condition is the same as one finds from analyzing paths of charged
particles,
which is a much simpler way to see the results.

The scaling symmetry discussed above shows up in the  form of the solutions
 near the horizon, and also in correlation lengths. Of course, this is not a
statistical
or quantum mechanical system
(though the fact that Hawking radiation makes it seem like one is intriguing)
so
to find correlation lengths we can't take an ensemble average. However,
we can compute the Greens
function for the wave equation, which tells how the classical field evolves in
response to a general source.  Quantum mechanically, the two point correlation
function is the
greens function. Next we show via the Green's function, that the influence
of a source dies off expontially for $Q<M$, whereas there is a long range tail
for $Q=M$.

Let a wave be incident on the black hole, with $\psi \sim {1\over
\sqrt{2\omega}}
 e^{-i\wep t -iky}$ as $y\rightarrow \infty $, with $\wep $ satisfying \capture
{}.
Then as $y\ra -\infty$, $\psi \ra \sqrt{{k\over \wep }}e^{-i \wep (t+y)}$. The
normalization
follows because the wronskian  of \wave \
is constant, and using the fact that the amplitude of the
captured wave is much greater than the amplitude of the scattered part. This is
only the
leading term in the asymptotic solution for $\psi$. More information will be
needed, so
next we find the leading nontrivial behavior of the wave near the horizon.

For $Q=M$  and for $y\ll -M $, the wave equation becomes
\eqn\horeq{ \pary ^2 \psi +\left[ \w-e ^2 -2e\w-e {M\over y}\right] \psi =0 .}
The inward propagating solution correct through order ${M\over y}$ is
\eqn\solone{\psi _1 =e^{-iS} e^{{eM\over 2\w-e y}} }
\eqn\sdef{ where \ S=\w-e y +eMln (-y) +{e^2 M^2 \over 2\w-e }{1\over y}.}
The outward propagating mode is given by
\eqn\soltwo{ \psi _2 =e^{iS} e^{-{eM\over 2\w-e  y }} }

For $Q<M$  and $\k y\ll -1$, the wave equation becomes
\eqn\horeqtwo{
\pary ^2 \psi +\left[ \wep ^2 +P_{\omega} e^{2\k y} \right] \psi = 0 }
where $P_{\omega} = 2e\pho \wep -2\k m^2 R_H$.
$P_{\omega}$ must be positive for the wave to actually get over the barrier,
by\capture. There
are two  (potentially) small parameters here, $\k$ which goes to zero in the
extremal
limit, and $\wep $ which we want small to be adding a small amount of mass to
the black hole. The regime of interest will be $\k \ll \wep $, to approach the
extremal black hole, and this is included in the $P_{\omega} >0$ range.

The inward propagating solution, correct through terms of order $e^{2\k y}$ is
\eqn\solthree{
\psi _3 =e^{-i\wep y} e^{-Ae^{2\k y}}  }
where $  A={P\over 4\wep ^2 +2\k ^2 }({2\wep
\over 2\k }+i)$.
For $\k \ll \wep $, $A\ra {e\pho \over 2\k} - {m^2 R_H \over 2\wep } +i({e\pho
\over  2\wep } -{2\k m^2  R_H \over 4\wep ^2 }).$ The outward propagating mode
is
the complex conjugate, \eqn\solfour{ \psi _4 =\psi _3 ^*  .}
As a consistency check, the wronskian of each of the above solutions $\psi _i
(y)$ is constant,
as it should be.

The (advanced) Greens function can now be constructed.  For $Q=M$ this is
\eqn\green{\eqalign{
G(t,y;t' ,y')=& -{1\over 2}\Theta (\dt ) \Theta (-\dy )\Theta (\dt +\dy )
e^{-ieMlog(y/y')} \left[ -{1\over 2}(1-ieM)({y'\over y }-1 ) \right.\cr
 &+\left. e^{-i{eM\over y'}(\dt +\dy )} e^{{1\over 2}(1-ieM)
({y'\over y} -1)} \right]\cr
&-{1\over 2}\Theta (\dt )\Theta (\dy )\Theta  (\dt -\dy )
e^{ieMlog ({y \over y' })} \left[ {1\over 2}(1-ieM)({y' \over y}-1)\right.\cr
&+\left.
e^{-i{eM \over y'}(\dt -\dy )}e^{-{1\over 2}(1-ieM)({y' \over y }-1)}\right]
 \cr }.}
For example, the field response to a delta function source at $y_o ,t_o$ is
\eqn\poke{
\psi (y,t) \approx e^{-ieMlog ({y\over y_o})}
[{1\over 4}(1-ieM)({y_o \over y} -1)
-{1\over 2}e^{-{1\over 2}(1-ieM)} e^{-i{eM\over y_o }(\dt + \dy )}
(1+{1\over 2}
{y_o \over y})] .}
The solution shows the scaling symmetry, and long-range correlations, i.e., the
effect
of the source falls off like $y^{-1}$. $G$ (or $\psi$) has a piece which is
independent of time, and
a piece which  goes to free oscillations at a frequency $eM/y_o $,
which depends on the location of the source point.

By contrast, for $Q<M$ there is no scaling symmetry,
and the influence of the source falls off exponentially fast, like $e^{2\k y}$.
The green's function has pieces which are oscillations at two frequencies, $\mu
\equiv e\pho e^{4\k y'}$,  and
$\k \gamma $, where $\gamma\equiv {m^2 R_H \over e \pho }$
The first depends on the location of the source point, and the second goes to
zero
in the extremal limit. Let $h(\dy ) =e^{2\k \dy } -1$. Then the greens function
for $Q<M$ is \eqn\greentwo{\eqalign {
G(y,t;y',t') =& \Theta (\dt ) \Theta (-\dy )\Theta (\dt +\dy ){1\over 2}
e^{-i{\mu \over 2\k}
h(\dy )} \left[ \mu e^{i \mu (\dt +\dy ) }e^{{1\over 2}(1-i\gamma )h (\dy )}
\right.\cr
&\left. -{\k\gamma \over \mu^2 }
e^{-i\k\gamma (\dt +\dy )} e^{-{\mu (1-i\gamma )\over
2\k\gamma }h(\dy )} \right] \cr
&+\Theta (\dt )\Theta (\dy )\Theta (\dt -\dy )
{1\over 2} e^{i {\mu \over 2\k }h(\dy ) }\left[ \mu
e^{i\mu (\dt -\dy )}e^{{1\over 2}(1+i\gamma )h(\dy ) } \right.\cr
&\left. -{\k\gamma \over \mu ^2 }e^{-i\k\gamma (\dt +\dy )}
e^{-{\mu (1+i\gamma )
\over 2\k \gamma }h(\dy ) }\right] \cr }.}
For  example, the field configuration at large negative values of $y$,
due to a point source, can be read off from the first
two lines, showing that $\psi$ approaches free oscillations exponentially fast.

\newsec{Critical Point Behavior?}\bk
For formation of a zero mass, neutral black hole,  the numerical
studies \chop ,\evans\ looked at how the mass
of the new black hole varied with the parameters of the incident wave. It was
found that this
behavior was non-analytic (and universal). When forming a charged  black hole,
the
relevant quantity may be fluctuations about the minimal area. In the present
context,
let us look at  infinitesimal changes in the area of an already existing black
hole, in the
limit where the black hole approaches extremal. (This is the analogue of the
approach to the
putative critical point.) For a black hole with general charge and mass, the
horizon radius is $R_H =M(1+\Delta )$ and the area is $4\pi R_H ^2 $. If
small amount of mass $\delta M$
and charge $\delta Q$ are added, the change in the horizon radius is
\eqn\rad{\delta R_H =\delta M +M\sqrt{\Delta ^2 +2{\delta M\over M} -2{Q\over
M}
{\delta Q\over M}} -M\Delta ,  }
where terms of order $\delta M ^2 ,\delta Q^2$ have been neglected.  For a wave
 carrying
$\delta Q=e$ to be captured, as discussed in \capture \ , it must have
\eqn\deltaM{\delta M\equiv \omega=e\pho +{m^2 \over e}(\Delta +\epsilon ) ,}
where $\epsilon \ra 0 $ to add the minimal possible mass.
Now there are two cases. If one fixes $\Delta$ and then considers $\delta M,
\delta Q
\ll \Delta$, then as  expected, one finds a formulae for the change in radius
which is linear in the perturbations to the mass and charge,
\eqn\radtwo{\delta R_H \approx e (\pho  -{Q\over M\Delta }) +m
{m\over e} (\Delta +\epsilon )
(1+{1 \over \Delta }) . }
On the  other hand, for the case of interest here, $m$ and $e$ are still small
compared to
$M$ and $Q$, but in addition, $\Delta \ra 0$. Precisely, for $\Delta \ll m^2
/(Me)$,
\eqn\radthree{\delta R_H \approx e+\crit . }
Therefore, in the limit where the change in the area is as small as possible
($\epsilon \ra
0$), the variation in the horizon area has non-analytic behavior, as the
extremal background
is approached.  Of course, \radthree \ could be written as linear in$\delta$,
where
$\delta =(1-{Q^2 \over M^2})^{1\over 4}$. Here $\Delta$  appeared to be a
natural
choice because the horizon area is polynomial in $\Delta$.
The suggestion in \radthree \ is that tuning through the background spacetimes
as
$\Delta \ra 0$ is like tuning the magnetic field to a critical value.

The First Law states that $\delta M={\k \over 8\pi}\delta A +\pho \delta Q$.
$\k$ and
$\pho$ play the roles of the temperature and an electric (or chemical)
potential, so
derivatives of $\delta R_H$ with respect to $\k ,\pho $ are also of interest.
Instead
of using $M$ and $ \Delta $ as the two independent variables to describe the
state
of the system,
we switch to the variables $\k ={2\Delta \over M(1+\Delta )^2}$ and $\pho
={\sqrt{1+
\Delta ^2}\over 1+\Delta }$. Then,
for example,
the analogue of the specific heat is
\eqn\spheat{  \k\left( {\partial \delta A\over \partial \k}\right) _{\pho} =
  4\pi \k R_H \left({\partial \delta R_H\over \partial \k}\right) _{\pho}
=-2\pi R_H \crit . }

\newsec{Discussion}
Consider the class of spacetimes consisting of a charged black hole,
parameterized by $M$ and $\Delta$, interacting with charged matter. We have
been looking at various properties
of this system, that suggests that the point $\Delta =0$ is like a critical
point.
To examine this, we move away from this point (look at $Q<M$ spacetimes ) and
probe
the system with charged test fields, to see the behavior as $\Delta \ra 0$.  It
was seen that fluctuations in the area of the black hole have non-analytic
behavior in this limit.
We showed that correlation lengths, defined in terms of the classical Greens
functions, are long range on the $\Delta =0$ background, and decay
exponentially
for spacetimes with $\Delta >0$. Further, we showed
that spherically symmetric packets of the test field evolve to configurations
which have a scaling symmetry
near the horizon, if and only if the background has $\Delta =0$. Near the
horizon,
the eigenmodes can be chosen such that each mode has a discrete
self-similiarity.
This is interesting, because the numerical studies of formation of neutral
black
holes showed that the collapsing field was discretely self-similiar, near the
critical
point. In the present study we don't know if the same is true; does an incident
wave
packet evolve into a special sum of the eigenmodes, such that the sum has a
discrete
self-similiarity. This is an interesting open question.

Fluctuations about zero mass  is a limiting case of fluctuations about
the minimal area (irreducible mass) , when the black hole is neutral.
This would suggest that a key feature to criticality is extremality. However,
there is
another possibility which is interesting to think about. A black hole with
$Q=M$
in a spacetime with positive cosmological constant is not extremal. However, it
does
have the property that the surface gravity of the black hole is equal in
magnitude
to the surface gravity of the deSitter Cauchy horizon, i.e., the two
temperatures
are the same. Geometrically, the spacetime geometry has an infinite throat near
the
horizon of the black hole, similiar to the geometry of the throat discussed
here.
Therefore, one might expect that the behavior of a charged test field would be
the same as in the present case. If this is true, then the key ingredient would
be equality of the Hawking temperature with the background.

{\bf Acknowledgements:} I would like to thank David Kastor for several useful
conversations,
and the Aspen Center for Physics for its hospitality. This work was supported
in
part by NSF grant NSF-THY-8714-684-A01.
\listrefs

\end